\documentstyle[epsf,aps,multicol]{revtex} 
\begin{document}
\draft

\title{A Crossed Sliding Luttinger Liquid Phase}

\author{Ranjan Mukhopadhyay, C.L. Kane, and T.C. Lubensky}
\address{Dept. of Physics, Univ. of Penn., Philadelphia, PA 19104}
\date{\today} \maketitle

\begin{abstract}
We study  a system of crossed spin-gapped and gapless Luttinger liquids. We 
establish the existence of a stable non-Fermi
liquid state with a finite-temperature,
long-wavelength, isotropic electric conductivity 
that diverges as a power law in temperature $T$ as $T\rightarrow 0$. 
This two-dimensional system has many properties characteristic 
of a true isotropic Luttinger
liquid, though at zero temperature it becomes anisotropic.  
This model can easily be extended to three dimensions. 
\end{abstract}
 

\begin{multicols}{2}

For over two decades a central theme in the study of correlated
electronic systems
has been the drive to understand and classify electronic states that 
do not conform to Landau's Fermi-liquid theory.  A clear example of such
``non-Fermi liquid"
physics occurs in one dimension (1D)\cite{voit}, where arbitrarily weak
interactions destroy the Fermi surface and 
invalidate the notion of independent quasiparticles at low energy.  
Away from charge-density-wave instabilities, the interacting 1D
electron gas
forms a Luttinger liquid in which the discontinuity in occupation at the Fermi
energy of a normal Fermi liquid is replaced by
a power-law singularity, and the low-lying excitations are bosonic
collective modes in which spin
and charge decouple.  

Following the discovery of high-temperature superconductivity, Anderson
suggested that the unusual normal-state
properties of the cuprates were the result of similar 
non-Fermi-liquid physics in two dimensions\cite{anderson1}.
However, the study of non-Fermi liquids in higher dimensions has proven
to be quite difficult.  Since the
Fermi liquid is stable for weak interactions, perturbative methods
fail\cite{randeria}.  
Moreover, generalizations of the bosonization technique to isotropic
systems in higher dimensions 
have indicated that Fermi liquid theory survives provided the
interactions are not pathologically 
long ranged\cite{marston}.  An alternative approach has been to study
anisotropic systems consisting 
of arrays of parallel weakly coupled 1D wires\cite{coupled}.
It has
recently been proposed\cite{efkl,vc} 
that for a range of interwire charge and current 
interactions, there is a {\it smectic-metal} (SM) phase in which
Josephson, charge- and spin-density-wave, and single-particle couplings are
irrelevant.  This phase is an anistropic sliding Luttinger liquid phase
whose transport properties exhibit power-law singularities like those of a
1D Luttinger liquid.  It is
the quantum analog of the sliding phases of coupled classical XY
models found by O'Hern {\it et al.} \cite{sliding,olt}.

We consider a square network of 1D wires formed by
coupling two perpendicular smectic metals\cite{castroneto}
 and show that it exhibits a new
 {\it crossed sliding Luttinger Liquid} (CSLL) phase. 
We establish a range of couplings for which both this
phase and the anisotropic two-dimensional smectic-metal phase 
from which it is constructed are stable with respect
to a large class of operators.
At finite temperature $T$, the CSLL phase is an
isotropic $2D$ Luttinger liquid with an isotropic long-wavelength
conductivity that diverges as a power-law in $T$ as $T\rightarrow 0$.  At
$T=0$, it is essentially two independent smectic metals.
This model could be realized in man-made strucures
constructed from quantum wires such as carbon nanotubes.
Extension of the model to a three-dimensional 
stack may be relevant to the stripe phases of the cuprates.
Based on neutron and x-ray scattering measurements, 
it has been 
suggested that spin-charge stripes in the adjacent CuO$_{2}$ plane
are orthogonal to each other
\cite{tranquada}.

The Lagrangian density describing the low-energy behavior of a
one-dimensional Luttinger liquid is
\begin{equation}
{\mathcal{L}}_{0}= {1\over 2}\kappa [ v^{-1}(\partial_{\tau} \phi)^2
 + v (\partial_{x} \phi )^2 ] ,
\label{lagrangian1}
\end{equation}
where $\phi$ is a bosonic field\cite{notation} and $\kappa$ and the sound
velocity, $v$, are non-universal functions of the coupling
constants. For repulsive interactions, $\kappa > 1$. 
The Lagrangian density in terms of  the dual phase variable $\theta$
has the same form as Eq.\ (\ref{lagrangian1}), but with $v$
replaced by $1/v$. For spin-$1/2$ fermions, 
the spin excitations could either be gapped or
gapless. In the spin-gap, Luther-Emery regime, the system can by
described by a single Luttinger liquid for charge. In the gapless
case, both spin and charge are dynamical degrees of freedom, and there
are two Luttinger parameters ($\kappa_c$, $\kappa_s$), and two velocities
($v_c$, $v_s$).

	Now consider a two-dimensional array of parallel quantum
wires.  To begin with, we consider the {\it{spin gapped}} case, so that 
the spin fluctuations on each wire are effectively frozen
out at low energies. It has been suggested that this
case might describe the stripe phases of
high-temperature superconductors \cite{emery}.
In general, we expect a generalized 
current-current interaction between the wires, which can be represented
by a Lagrangian density of the form
\begin{equation}
{\mathcal{L}}_{\rm int}={1 \over 2} \sum_{n,n',\mu}
j_{\mu,n}(x,\tau) \tilde{W}_{\mu}(n-n')j_{\mu,n'}(x,\tau) ,
\label{lagrangian2}
\end{equation}
where $j_{\mu ,n}=(\rho_{n}(x,\tau),J_{n}(x,\tau))$ with
$\rho_{n}=\partial_{x}\phi_{n}(x,\tau)$ the density and
$J_{n}=\partial_{\tau}\phi_{n}(x,\tau)$ the current on the $n$th
wire. This interaction is marginal and should be included in the
fixed-point action. It is invariant under the ``sliding"
transformations $\phi_{n} \rightarrow \phi_{n} + \alpha_{n}$ and
$\theta_{n} \rightarrow \theta_{n} + \alpha^{\prime}_{n}$. Equations
(\ref{lagrangian1}) and (\ref{lagrangian2}) define the fixed-point
action of the smectic-metal phase\cite{efkl}, which can be written
in Fourier space as
\begin{eqnarray}
S &=&\sum_{Q}{1 \over 2}\{W_{0}(q_{\perp})\omega^{2} +
W_{1}(q_{\perp})q_{\parallel}^{2}\} |\phi(Q)|^{2} , 
%\nonumber \\ &=& \sum_{Q}{1
%\over 2}\left\{{\omega^{2} \over {W_{1}(Q)}} + {{k^{2}} \over
%{W_{0}(Q)}} \right\} |\theta(Q)|^{2}
\end{eqnarray}
where $Q=(\omega, q_{\parallel}, q_{\perp})$, with $q_{\parallel}$
the momentum along the chain and $q_{\perp}$ perpendicular to
the chains. 
            
     We can study perturbatively the relevance of various operators to
ascertain the stability of the smectic-metal (SM) phase. Due to the spin
gap, single-particle hopping between chains is irrelevant, and the
only inter-chain interactions involving only pairs of chains
that could become relevant are 
the Josephson (SC) and CDW couplings, whose respective Hamiltonian
densities are 
\begin{eqnarray}
{\mathcal{H}}_{{\mathrm SC},n} &=& \sum_{i} {\mathcal{J}}_{n} \cos[\sqrt{2
\pi}(\theta_{i} - \theta_{i+n})], \nonumber \\ {\mathcal{H}}_{{\mathrm CDW},n} &=&
\sum_{i} {\mathcal{V}}_{n} \cos[\sqrt{2 \pi}(\phi_{i} - \phi_{i+n})] ,
\end{eqnarray}
where ${\mathcal{J}}_n$ are the inter-chain Josephson couplings and
${\mathcal{V}}_{n}$ the inter-chain particle-hole (CDW) interactions.
The scaling dimensions of ${\mathcal{H}}_{{\mathrm SC},n}$ and
${\mathcal{H}}_{{\mathrm CDW},n}$ are, respectiveley, 
\begin{eqnarray}
\Delta_{{\mathrm SC},n}&=&\int_{- \pi}^{\pi} {{d q_{\perp}} \over {2 \pi}}
(1 - \cos{n q_{\perp}}) \kappa(q_{\perp}), \nonumber \\
\Delta_{{\mathrm CDW},n}&=&\int_{- \pi}^{\pi} {{d q_{\perp}} \over {2
\pi}} {{(1 - \cos{n q_{\perp}})} \over {\kappa(q_{\perp})}} ,
\label{eq: stability}
\end{eqnarray}
where $\kappa(q_{\perp}) = \sqrt{W_0(q_{\perp}) W_1 (q_{\perp})}$.
For the smectic-metal phase to be stable, these perturbations should be
irrelevant, which implies 
\begin{equation}
\Delta_{{\mathrm CDW},n} > 2, \ \ \  \Delta_{{\mathrm SC},n'} > 2
\end{equation}
%$\Delta_{{\mathrm CDW},n} > 2$ and $\Delta_{{\mathrm SC},n'} > 2$
for all $n$ and $n'$. In addition to the pairwise operators of
Eq.~(4), there are multiwire operators of the form 
${\mathcal H}_{{\rm CDW},\{\sigma_{n}\}} = \sum_{i}
{\mathcal T}(\sigma_{n})\cos[\sqrt{2 \pi}(\sum_{n}\sigma_{n}
\theta_{i+n})]$ where the $\sigma_{n}^{'}$s are integers 
satisfying $\sum \sigma_{n}=0$. The overall strengths of these
interactions measured by ${\mathcal T}(\sigma_{n})$ are
much smaller than those of ${\mathcal H}_{{\rm CDW},n}$,
and they become important only at very small temperatures even if 
they are relevant. We will therefore ignore them in this article,
delaying a more complete study of their effects to a future
publication. 
%Note, however, that
%we have included the effects of higher order superconducting
%terms in our stability analysis.

To explore the regions of stability of the SM phase, we follow
refs.\cite{vc,olt} and take
\begin{equation}
\kappa(q_{\perp})= K[1 + \lambda_{1}\cos(q_{\perp}) + \lambda_{2}
\cos(2 q_{\perp})] .
\label{kappa}
\end{equation}
We define 
$\Delta_{{\mathrm SC},n}=a_n K$ and $\Delta_{{\mathrm CDW},n}= b_n /K$,
where $a_{1}=(1 - \lambda_{1}/2)$, $a_{2}=(1-\lambda_{2}/2)$ and
$a_{n}=1$ for $n \neq 1,2$. The SM phase becomes unstable
to inter-chain Josephson couplings for $K$ less than
$K_{\mathrm{SC}}={\mathrm{max}}_{n} (2/a_{n})$ and unstable to 
inter-chain CDW
interactions for $K$ greater than $K_{\mathrm{CDW}}={\mathrm
{min}}_{n} (b_{n}/2)$.
Thus the smectic metal phase is stable with respect to
pairwise interactions over a window of $K$,
$K_{\mathrm {SC}}< K < K_{\mathrm CDW}$, provided
\begin{equation}
\beta \equiv {K_{\mathrm{CDW}} \over K_{\mathrm {SC}}}=
 \left. {{a_n b_m} \over 4} \right|_
{{\mathrm{min. wrt.}} m \& n}> 1.
\end{equation}
If $\beta < 1$, the system goes directly from a 2D superconducting
(SC) phase to a CDW crystal as $K$ is increased, without passing
through the SM phase. 
$\kappa(q_{\perp})$
reaches a minimum, $\kappa (k_0 ) \equiv \Delta K$ at some
$q_{\perp}=k_{0}$.
The SM phase is stabilised at small
$\Delta$, that is, when the system is
close to a CDW instability in which there is a
periodic modulation of charge on different wires. Setting $\Delta = 10^{-5}$,
we plot $\beta$ as a function of $k_{0}$. 
%In this ansatz, while
%$\lambda_{2}$ is always positive,  $\lambda_{1}$ has the same sign
%as $(-\cos k_{0})$. 
 We note that there are regions of stability of the smectic
phase with respect to ${\mathcal H}_{\rm SC}$ and
${\mathcal H}_{\rm CDW}$,
for positive as well as negative values of $\lambda_{1}$.

\begin{figure}
\par\columnwidth20.5pc
\hsize\columnwidth\global\linewidth\columnwidth
\displaywidth\columnwidth
\epsfxsize=3.0truein
\centerline{\epsfbox{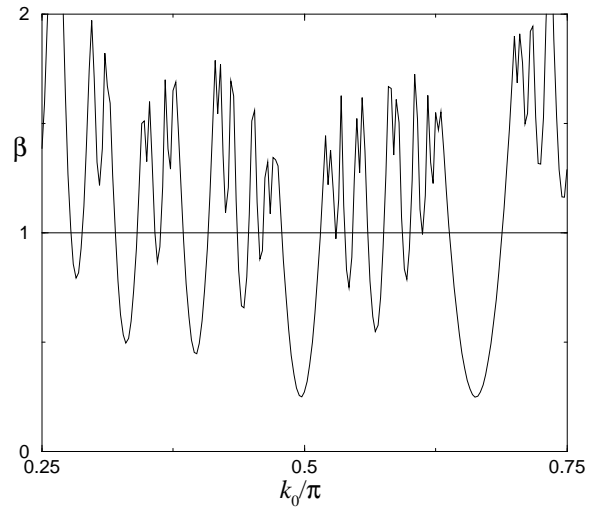}}
\vspace*{.1in}
\caption{ Plot of $\beta \equiv K_{CDW}/K_{SC}$ as a function of
of $k_{0}/\pi$. For $\beta > 1$ there exists a region of $K$ over
which the non-Fermi liquid phase is stable to all pairwise interactions.}
\label{fig:EF1}
\end{figure}

% For small 
%$\Delta$, $b_{n}$'s are approximately given by
%\begin{equation}
%b_n \simeq {{1 - \cos(n k_{0})e^{-n \sqrt{\Delta/C}}} \over \sqrt{C \Delta}}
%\end{equation}
%where $C \equiv f''(k_{0})/2 = 2 \lambda_{2} \sin^{2}k_{0}$.
% We note that there are regions of stable smectic
%phase for positive as well as negative $\lambda_{1}$.
% If $\beta > 1$, as
%$K$ is increased at constant $\gamma_{1}$ and $\gamma_{2}$, we go from
%the superconducting state, through the smectic metal state, to a
%charge density wave crystalline phase.

We consider
next a square grid of wires, again for the spin-gapped case.  There are
two  arrays of quantum wires, the $X$- and $Y$- arrays running,
respectivley, parallel to the $x$- and $y$-directions. 
Each wire now sees a periodic one-electron
potential from the array of wires crossing it 
%For simplicity we assume that it is commensurate with with bands in the wire. 
that leads to a new band structure with new band
gaps. We assume that the Fermi surface is between gaps so that the
wires would be conductors in the absence of further interactions. By
removing degrees of freedom with wavelengths smaller than the inverse wire
separation, we obtain a new effective theory whose form is identical to
the theory before the periodic potential was introduced. Thus, in the
absence of two-particle interactions between crossed arrays, the system
could be in a phase consisting of two crossed, non-interacting 
smectic-metal states.

 We will now demonstrate that the sliding phase 
in a crossed grid is stable if the sliding phase in 
the component planar arrays is stable.
In addition to the interwire couplings
within each array, we need to consider Coulomb interactions between 
wires on the X-array and wires on the Y-array. These 
inter-array couplings are marginal and should be included
in the fixed point. They do not, however, change the
dimensions of the operators, except by renormalizing
$\kappa(q_{\perp})$. For a stable sliding phase, 
additional interactions between the two arrays, such as 
the Josephson and CDW couplings, have to be irrelevant.
We will show that it is possible to tune $\kappa(q_{\perp})$
such that this is indeed the case.

The Coulomb interactions between electrons on
intersecting wires gives rise to a term in the Hamiltonian
of the form $V^{c}_{m,n}(x,y)\rho_{x,m}(x)\rho_{y,n}(y)$, 
where $\rho_{x,m}(x)$ [$\rho_{y,m}(y)$] is the electron density on the
$m$th wire on the $X$($Y$)-array at position
$x$ ($y$). We expect $V^{c}_{m,n}(x,y)$ to have the form $V^{c}(x - na,
y - ma)$, where $a$ is the distance between parallel wires. 
If all parameters for the $X$ and $Y$-arrays
are the same, the crossed-grid action as a functional
of the $\theta$ and $\phi$ variabless can be written as
\begin{eqnarray}
    S & = &{1 \over 2} \int {{d\omega dq_{x} dq_{y}}\over{(2 \pi)^{3}}}
   [V^{\theta}(q_{y}) q_{x}^{2} |\theta_{x}|^{2}
   + V^{\theta}(k_{x}) q_{y}^{2}|\theta_{y}|^{2}
 \nonumber \\
 & & + V^{\phi}(q_y) q_{x}^{2}|\phi_{x}|^{2}
+  V^{\phi}(q_x) q_{y}^{2}|\phi_{y}|^{2} \nonumber \\
& &+ V^{C}(q_{x},q_{y})q_x q_y \{\phi_{x} \phi_{y}^{*}
 + {\mathrm{c.c.}}\} \nonumber \\
& & -i \omega q_{x}\{ \theta_{x}^{*} \phi_{x} + {\mathrm{c.c.}}\}
 -i \omega q_{y}\{ \theta_{y}^{*} \phi_{y} + {\mathrm{c.c.}}\}]
\end{eqnarray}	
with obvious definitions for $\phi_x = \phi_x(\omega, q_x, q_y), \phi_y,
\theta_x$, and $\theta_y$.
It should be noted that this is an effective theory with
 $-{\pi \over a} < q_{x},q_{y} < {\pi \over a}$. 
%Also since $|k_{x}|, |k_{y}|$
%are smaller than ${{\pi} \over a}$,
%we could assume $V^{C}(k_x,k_y)$ is a constant. 
Integrating out the $\phi$ variables, we are left with an effective 
action
\begin{eqnarray}
  S_{\theta}& = & {1 \over 2} \int {{d\omega dq_{x} dq_{y}}\over{(2\pi)^{3}}}
\left[{1 \over \kappa_{x}({\mathbf{q}})}\left(v_{x}({\mathbf{q}}) q_{x}^{2}
+ {{\omega^{2}}\over{v_{x}({\mathbf{q}})}}\right)|\theta_{x}|^{2}
\right. \nonumber\\
& &+ {1 \over {\kappa_{y}({\mathbf{q}})}} \left(v_{y}({\mathbf{q}})
q_{y}^{2}+ 
{{\omega^{2}}\over{v_{y}({\mathbf{q}})}} \right)|\theta_{y}|^{2}
\nonumber \\
& &+ \left. V^{c}_{R}({\mathbf{q}}) \omega^{2} \{\theta_{x} \theta_{y}^{*}
 + {\mathrm{c.c.}}\}\right], \label{eq:action} 
\end{eqnarray}
where $\kappa_{x}({\mathbf q})=\sqrt{{\gamma({\mathbf q})} \over
{V^{\phi}(q_{x})V^{\theta}(q_{y})}},$ \ \
$v_{x}({\mathbf{q}})=\sqrt{{V^{\theta}(q_{y})\gamma({\mathbf{q}})}
\over{V^{\phi}(q_{x})}}$ \\
with $\gamma({\mathbf{q}})
= V^{\phi}(q_{x})V^{\phi}(q_{y}) - (V^{c}({\mathbf q}))^{2},$
and $\kappa_{y}({\mathbf q})=\kappa_{x}(P{\mathbf K})$; 
$v_{y}({\mathbf{q}})=v_{x}(P{\mathbf q})$ where 
$P {\mathbf{q}}=P(q_{x},q_{y})=(q_{y},q_{x})$. Correlation functions
for $\theta_{x}$ and $\theta_{y}$ can be calculated directly
from Eq. (\ref{eq:action}). $\theta_{x}$-$\theta_{y}$ cross-correlations
are non-singular, whereas, $\theta_{x}$-$\theta_{x}$ and 
$\theta_{y}$-$\theta_{y}$
correlations have singular parts with exactly the same functional
forms as they have in the absence of coupling between layers,
but with the $\kappa(q)$ function in expressions for the scaling 
exponents replaced by
\begin{equation}
 \kappa(q_{\perp})=\kappa_{x}(0,q_{\perp})=\kappa_{y}(q_{\perp},0) .
\end{equation}
Thus, other than renormalizing
$\kappa(q)$, the coupling $V^{c}_{m,n}$ between the two arrays leaves the dimensions
of all operators {\it unchanged}.
Equations (10) and (11) define a 2D non-Fermi liquid with
scaling properties to be discussed below. 

First, however, we must verify that it is possible to
choose potentials so that this 2D non-Fermi liquid is stable with respect to
perturbations. All pairwise couplings within a given array,
i.e. ${\mathcal{H}}^{X}_{{\mathrm SC},n}$, ${\mathcal{H}}^{X}_
{{\mathrm CDW},n}$, ${\mathcal{H}}^{Y}_{{\mathrm SC},n}$ and 
${\mathcal{H}}^{Y}_{{\mathrm CDW},n}$ defined as obvious generalizations 
of Eq.~(4), can be rendered irrelevant by choosing $\kappa(q_{\perp})$ 
as in the case of an individual array. We must 
also consider Josephson and CDW couplings between the two arrays,
which operate at the points of crossing $(x,y) = (na, ma)$ of wire
$m$ in the $X$-array and wire $n$ of the $Y$-array and,
respectively, take the form
\begin{eqnarray}
{\mathcal{H}}^{XY}_{\mathrm{SC}}&=&{\mathcal{J}}^{XY} \cos[\sqrt{2
\pi} (\theta_{x,m}(na) - \theta_{y,n}(ma))] \nonumber \\
H^{XY}_{\mathrm{CDW}} &=& {\mathcal{V}}^{XY} \cos[\sqrt{2 \pi}
(\phi_{x,m}(na) - \phi_{y,n}(ma))].
\end{eqnarray}
The dimensions of these operators are, respectively,
\begin{eqnarray}
\Delta_{\mathrm{SC},\infty} &\equiv& \int^{\pi}_{-\pi} {dq \over 2 \pi} \kappa(q)
= K \nonumber \\
\Delta_{\mathrm{CDW},\infty} &\equiv& \int^{\pi}_{-\pi} {dq \over 2 \pi} {1
\over \kappa(q)} \simeq {1 \over K} {1 \over \sqrt{C \Delta}} ,
\end{eqnarray}
where we assume that $\kappa(q)$ has the form given by Eq. (7),
 $\Delta$ is defined as before, and $C \equiv \kappa''(k_{0})/2K$.
If $\kappa$
is chosen such that Eq.\ (\ref{kappa}) is satisfied for each array, then
${\mathcal{H}}^{XY}_{\mathrm{SC}}$ and ${\mathcal{H}}^{XY}_{\mathrm{CDW}}$
are automatically irrelevant. We do not need any further fine tuning
of $\kappa$ to get this 2D sliding Luttinger liquid phase, and there is a
stable CSLL phase. 

We now investigate the transport properties of the CSLL phase. 
The conductivities
of an array of parallel wires has been considered by Emery 
{\it {et. al}} \cite{efkl}. In the presence of impurities,
the resistivity along the
wires, $\rho_{\parallel}$, vanishes as $T^{\alpha_{\parallel}}$\cite{lp}, with
$\alpha_{\parallel}= \Delta_{{\mathrm{CDW}},\infty} - 2$.
The perpendicular conductivity, $\sigma_{\perp}$,
goes as $T^{\alpha_{\perp}}$ with
$\alpha_{\perp}=2\Delta_{\mathrm{SC}}-3$,
where $\Delta_{\mathrm{SC}}$
is the minimum of $\Delta_{\mathrm{SC},1}$ and $\Delta_{\mathrm{SC},2}$. 
The conductance, $\sigma_c$, arising from the Josephson coupling at the
contact beteen the crossed wires satisfies 
$\sigma_{c} \sim T^{\alpha_{c}}$, 
where $\alpha_{c}=2\Delta_{\mathrm{SC},\infty}-3$.

\begin{figure}
\par\columnwidth20.5pc
\hsize\columnwidth\global\linewidth\columnwidth
\displaywidth\columnwidth
\epsfxsize=3.0truein
\centerline{\epsfbox{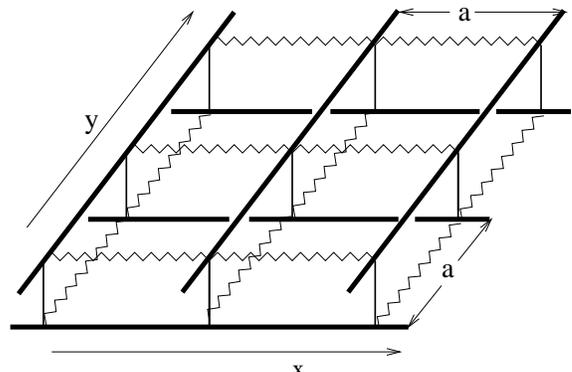}}
\caption{A schematic depiction of the 2D non-Fermi liquid as a resistor
network, with two parallel arrays of wire running along
the $x$ and $y$-axes, with nodes in the $z$ direction}
\label{fig:EF2}
\end{figure}

Thus we can model our 2D non-Fermi liquid as the resistor
network depicted in Fig.\ (2) with nodes at the vertical
Josephson junctions between the arrays at $r_{mn}$. The nodes
of the $X$($Y$)-array are connected by nearest neighbor
resistors with conductances $\sigma_{\parallel}=\rho_{\parallel}^{-1}$
if they are parallel to the $x$($y$)-axis and $\sigma_{\perp}$
if they are perpendicular to the $x$-axis($y$-axis). Nearest neighbor
nodes of the $X$ and $Y$-arrays are connected by resistors
of conductance $\sigma_{c}$. In the continuum limit, the 2D
current densities in the plane of the $\alpha$ grids ($\alpha=X,Y$) 	
is $J_{i}^{\alpha}=\sigma^{\alpha}_{ij}E_{j}$ where
$\sigma_{ij}^{X}=\sigma_{\parallel}e_{xi}e_{xj} + \sigma_{\perp}
e_{yi}e_{yj}$ and $\sigma^{Y}_{ij}=\sigma_{\perp}e_{xi}e_{xj}
+\sigma_{\parallel}e_{yi}e_{yj}$ and ${\mathbf E}$ is the 
in-plane electric field. The current per unit area
passing between the planes
is $J_{n}=(\sigma_{c}/a^{2})(V^{X}-V^{Y})$ where $V$ is the
local voltage. In this limit, the local voltages satisfy
\begin{eqnarray}
-\sigma_{ij}^{X}\partial_{i}\partial_{j}V^{X} + {\sigma_{c} \over a^{2}}
(V^{X}-V^{Y}) &=& {\mathcal T}^{X} \nonumber\\ 
-\sigma_{ij}^{Y}\partial_{i}\partial_{j}V^{Y} - {\sigma_{c} \over a^{2}}
(V^{X}-V^{Y}) &=& {\mathcal T}^{Y},
\end{eqnarray}
%the equation
%\begin{equation}
%\left( \begin{array}{cc}
%-\sigma_{ij}^{X}\partial_{i}\partial_{j} + {\sigma_{c} \over a^{2}}
%& -\sigma_{c}/a^2 \\
%-\sigma_{c}/a^{2} & -\sigma_{ij}^{Y}\partial_{i}\partial_{j}
%+ {\sigma_{c} \over a^{2}}
%\end{array} \right)
%\left( \begin{array}{c}
%V^{X} \\ V^{Y}
%\end{array} \right) =
%\left( \begin{array}{c}
%{\mathcal{T}}^{X} \\ {\mathcal{T}}^{Y} 
%\end{array} \right)
%\end{equation}
where ${\mathcal{T}}^{X}$ and ${\mathcal{T}}^{Y}$ are current
densities (current/area) injected, respectively, into the $X$ and
$Y$-grids. If no currents are injected, then this equation is 
solved by $V^{X}=V^{Y}=-{\mathbf E}\cdot{\mathbf x}$ to produce a
total in-planar curent density
\begin{equation}
J_{i}\equiv J^{X}_{i}+J^{Y}_{i}=(\sigma^{X}_{ij} + \sigma^{Y}_{ij})E_{j}
=(\sigma_{\parallel}+\sigma_{\perp})E_{i}.
\end{equation}  
Thus under a uniform electric field, the double layer behaves like 
an isotropic 2D material with in-plane
conductivity $\sigma=\sigma_{\parallel}+\sigma_{\perp}
\simeq \sigma_{\parallel}$, or equivalently with an isotropic
resistivity that vanishes as $\rho_{\parallel} \sim
T^{\alpha_{\parallel}}$.  If currents are spatially nonuniform, as they
are, for example, when current is inserted at one point and extracted
from another, there is a crossover from isotropic to anisotropic behavior
at length scales less than
$l= a\sqrt{{\sigma_{\parallel}+ \sigma_{\perp}} \over \sigma_{c}}
\sim T^{-(\alpha_{\parallel}+\alpha_{c})/2}$
that diverges as $T\rightarrow 0$.  
%At $T=0$, current can only be carried
%along the wires: the resistance between wires in a grid or between grids is
%infinite.

This two-layer CSLL model can, quite simply, be
extended to three dimensions by stacking alternate arrays in the third
direction. It is still possible, all be it more difficult,
for a stable CSLL phase to exist. This phase is characterized 
by an isotropic in-plane conductivity 
$\sigma_{\parallel} \sim T^{-\alpha_{\parallel}}$
and a conductivity $\sigma_{c}/a \sim T^{\alpha_{c}}$
in the direction perpendicular to the planes of the wires.
 Thus the conductivity along the planes
is much larger than the perpendicular conductivity. 
%    HALL CONDUCTIVITY?

The extension of the above analysis to a system of coupled Luttinger
liquids where both charge and spin excitations are {\it {gapless}} is
straightforward.   On each wire, there is a Luttinger liquid for charge and
for spin.   To maintain gapless 
Luttinger liquids and $SU(2)$ symmetry, we require that the spin 
degrees on each wire be represented by a Lagrangian of the form Eq.\
(\ref{lagrangian2})
with $\kappa_{s}=1$ and that at the fixed point there be no spin coupling 
between the wires. The fixed point for the charge degrees of freedom has
the same form as for the gapped case. However, now, single-particle
tunnelling as well as Josephson and CDW couplings may be relevant. 
The phase diagram is quite complicated and will be discussed in a 
future publication. There {\it is}, however, 
a small but finite region of phase space where the sliding phase is stable.

    In conclusion, we have demonstrated the existence
of a non-Fermi metallic phase in two dimensions
that maintains spin-charge separation and
is stable to all pairwise potentials. This is a remarkable
phase, which
could be identified as a two-dimensional Luttinger liquid. 
We delay to a future publication the investigation of the
stability of this phase with respect to all multiwire interactions.

%Whether its existence is of relevance to the normal
%conducting phases of the cuprates  needs to be explored.
%Our model differs from earlier work on network of
%crossed wires\cite{castroneto}, in that we have included a more complete set
%of operators in our model which drastically alter the phase
%diagram, with novel phases such as the CSLL phase, which
%have not been postulated before. 

  RM and TCL acknowledge support from the National Science
Foundation under grant DMR97-30405.

\end{multicols}


\begin{thebibliography}{10}
\bibitem{notation} There has been some inconsistency in the
literature regarding the definitions of $\theta$ and $\phi$.
In this paper, we follow the notation of\cite{efkl}, and
define $\theta$ as the phase variable, and $\phi$ as the
conjugate density variable.
\end{thebibliography}
\end{document}